\documentstyle[pra,twocolumn,aps,epsfig]{revtex}

\begin{document}

\draft

\title{Probing the statistical properties of Bose-Einstein condensates 
with light}

\author{Zbigniew Idziaszek$^{1,2}$, Kazimierz Rz\c a\.zewski$^1$, 
and Maciej Lewenstein$^2$}

\address{(1) Center for Theoretical Physics and College of Science\\ 
Polish Academy of Sciences, Al. Lotnik{\'o}w 32/46, 02-668 Warsaw}

\address{(2) Institut f\"ur Theoretische Physik, Universit\"at
Hannover, D-30163 Hannover, Germany}

\maketitle

\begin{abstract}
A scattering of short, weak, nonresonant laser pulses on a Bose-Einstein 
condensate is proposed as a tool for studying its statistical properties.
We show in particular that the variance of the number of scattered photons may distinguish between the Poisson and microcanonical statistics. 
\end{abstract}


%\narrowtext
\section{Introduction}

The mean field approach to describe properties of Bose--Einstein condestates 
(BEC) in weakly interacting atomic gases \cite{jila,mit,rice,edwards} 
has proved to be immensly succesful \cite{revmod}. However, 
quantum statistical properties of the condensate, 
and in particular higher order correlations in BEC have been a subject of
considerable interest in the recent years. In the series of papers the 
fluctations of the number of condensed atoms of the ideal Bose gas 
were calculated in micro- and canonical ensembles \cite{ideal}.  
The corrections due to the weak interatimic interactions were considered by 
several authors, but different models give different 
results, so that the impact of collisions on the statistics remains unclear 
\cite{nonide}. 
The possible reason for these diffuculties is the ambiguity in defining 
condensed fraction. Similar ambiguities lead to difficulties 
in determinations of the modification of the critical 
temperature for weakly interacting 
gases \cite{temper}. 

While the mean number of condensed atoms $N_0$
 as a function of temperature has been 
measured in several laboratories \cite{mitjil}, 
from the experimental point of view very little is known about 
fluctuations of BEC. The main source of information concerning higher 
order correlations comes from the studies of the depletion of 
BEC due to  inelastic two-body and three-body collisions \cite{kagan}.
In this way the mean value of $N_0^2$ and $N_0^3$, have been estimated
\cite{jila23}. The experimental result have unequivocally ruled out
the thermal fluctations in the condensate. Precision of those measurements is, 
however, unable to distinguish between sub- and super-Poissonian, but 
normal fluctuations.

In fact different models of the interacting atoms condensate coexist in 
literature. In particular, the classic Bogolyubov approximation \cite{Bogol} 
tends to 
favour the Poisson statistics of the condensate. It would be desirable to 
discriminate between the models with the help of a clean experiment.  

This paper 
describes such a proposal.  We propose to use a scattering of a short 
weak non-resonant laser pulse as a means of probing the BEC statistics. 
The problem of light scattering on the condensate has been thoroughly
studied in the recent years \cite{light}.
Most of these papers, however, concerned the signatures of BEC transition in
scattering, rather statistics of the condensate. The references most closely
related to the present paper are papers by You {\it et al.} \cite{You,Youferm} 
on scattering of short laser pulse on bosonic and fermionic clouds. These
papers use very similar approximations to solve 
time-dependence of the system, but both use grand canonical ensemble, and do
not address the problem of the condensate fluctuations.  

The paper is organized as follows. In section II we present the 
theoretical model, and solve it for the photon operators of the 
scattered field. In Section III we analyze the mean number of 
scattered photons,
whereas in Section IV its  variance in function of the assumed
statistical properties of the condensate, and physical 
parameters such as temperature and scattering angle. We compare the result 
for the thermally fluctuating gas, for the coherent state, 
for the ideal Bose gas described by the 
microcanonical ensemble and for the nonfluctuating gas. While
 the mean number of scattered photons can 
only discriminate between the thermal 
and the considered non-thermal 
condensate states, the variance can further 
distinguish the coherent state from the last two models.   
In Section V we present some conclusions.

\section{The model}

Our method of exploring the statistical properties of the condensate is based 
on the scattering
of the series of short light pulses. The fluctuations in a given 
condensate occur at the time scale given by the interatomic 
collisions, thus to probe the fluctuations the time delay 
between consecutive pulses should be of the order of miliseconds.
 The distribution of the
number of photons scattered into a given solid angle should be measured. 
Out of the distribution of the number of
scattered photons we may compute the mean and its variance. 
We assume that the pulse of light is weak and far detuned in order to avoid 
heating the condensate during the interaction. The
pulse of light should be also sufficiently short in time.
It should satisfy two conditions: 1. $\omega_{max} t \ll 1$
where  $\hbar \omega_{max}$ is the energy of the highest occupied level in
the trap, since we are going to ignore factors $e^{i \omega_n t}$, 
2. $t \ll \gamma^{-1}$ or the lifetime of the transition, since we are going
to ignore the spontaneous emission. Note, that the conditions for 
the proposed experiment are complementary to 
the well known method of phase contrast used for nondestructive 
imaging of the condensate \cite{imaging}. In fact we are interested in 
the regime in which the (small) condensate does not lead to significant 
refraction,
while the scattered light carries then more 
direct information about fluctuations.

The total Hamiltonian consists of the following parts
\begin{equation}
\label{HamTot}
\hat{H} = \hat{H}_{a} + \hat{H}_{al} + \hat{H}_{af} + \hat{H}_{f}, 
\end{equation}
The term $\hat{H}_{a}$ is the atomic Hamiltonian. In the second
quantization formalism it reads
\begin{equation}
\label{HamA}
\hat{H}_{a} = \sum_{\vec{n}} \hbar \omega_{\vec{n}}^{g}
\hat{g}_{\vec{n}}^{\dag} \hat{g}_{\vec{n}} + 
\sum_{\vec{m}} \hbar (\omega_{\vec{m}}^{e} + \omega_{0})
\hat{e}_{\vec{m}}^{\dag} \hat{e}_{\vec{m}},
\end{equation}
where $\hbar \omega_{\vec{n}}^{g}$ and $\hbar \omega_{\vec{m}}^{e}$ are
the energies of the CM motion of atoms, in the ground and the
excited electronic state respectively. The $\omega_{0}$ is the atomic 
resonance frequency.
The operators $\hat{g}_{\vec{n}}$ ($\hat{e}_{\vec{m}}$) are annihilation
operators of atoms in ground (excited) electronic state respectively. They fulfill
bosonic commutation relations
$[\hat{g}_{\vec{n}},\hat{g}_{\vec{m}}^{\dag}]=\delta_{\vec{n},\vec{m}}$,
$[\hat{e}_{\vec{n}},\hat{e}_{\vec{m}}^{\dag}]=\delta_{\vec{n},\vec{m}}$.
We use vector indices $\vec{n}$ and $\vec{m}$ because we consider 
a three dimensional trap.

The term $\hat{H}_{f}$ in the Hamiltonian is the energy of the
electromagnetic (e-m) field:
\begin{equation}
\label{HamF}  
\hat{H}_{f} = \sum_{\lambda} \int d^{3}k \
\hbar \omega_{\vec{k}} 
\hat{a}_{\vec{k} \lambda}^{\dag} \hat{a}_{\vec{k} \lambda}, 
\end{equation}
where $\hat{a}_{\vec{k} \lambda}$ ($\hat{a}_{\vec{k} \lambda}^{\dag}$) are
annihilation and creation operators of photons with wave-vector $\vec{k}$
(and frequency $\omega_{\vec{k}} = k c $) and polarization $\lambda$
respectively.

The interaction of the atoms with the laser light is described by
the $\hat{H}_{al}$ term:
\begin{equation}
\label{HamAL}  
\hat{H}_{al} = \frac{\hbar \Omega}{2} \sum_{\vec{n},\vec{m}} 
\langle \vec{n},g | e^{-i \vec{k}_{L} \vec{r}} | \vec{m},e \rangle
e^{i \omega_{L} t} \hat{g}_{\vec{n}}^{\dag} \hat{e}_{\vec{m}} + h. c.
\end{equation}
where $\omega_{L}$ is the laser frequency, $\Omega = 2 \vec{E} \vec{d} /
\hbar$ is the Rabi frequency with the transition dipole moment $\vec{d}$ 
and amplitude of the electric field of the laser $\vec{E}$. In the above 
equation Franck-Condon factors 
$\langle \vec{n},g | e^{-i \vec{k}_{L} \vec{r}} | \vec{m},e \rangle$ appear.
They are proportional to the amplitude of the
transition between two states of the trap induced by absorbtion or emission
of a photon. The trapping potential felt by excited and ground-state atoms 
may in general be different, and therefore their eigenstates may also 
be different. 
The labels $e$ and $g$ distinguish these two sets of states. However, in
later derivations we will assume the same potential for all atoms, and we
will use notation $| \vec{n} \rangle$ for the single atom trap states.

The atom-field interaction part of the Hamiltonian has the form:
\begin{eqnarray}
\label{HamAF}
\hat{H}_{af} & = & i \sum_{\lambda} \int d^{3}k \
\sqrt{\frac{\hbar \omega_{\vec{k}}}{2 \varepsilon_{0} (2\pi)^3}} 
(\vec{d} \cdot \vec{\epsilon}_{\vec{k} \lambda}) 
\hat{a}_{\vec{k} \lambda}^{\dag} \times \nonumber \\
& & \times \sum_{\vec{n} \vec{m}}
\langle \vec{n},g | e^{-i \vec{k}_{L} \vec{r}} | \vec{m},e \rangle 
\hat{g}_{\vec{n}}^{\dag} \hat{e}_{\vec{m}} + h. c.
\end{eqnarray}    

The Heisenberg equations of motion for the atomic operators 
$\hat{g}_{\vec{n}}$, $\hat{e}_{\vec{m}}$ can be solved in
the following approximation. The atoms are driven only by laser field 
dominating over vacuum modes. In our approach we neglect the back action of
atoms on the driving light and the influence of the vacuum modes. This last
approximation is just a neglect of the spontaneous emission. It is justified
if the duration of the pulse is short compared to the spontaneous emission
life time. The laser light, as one may notice from 
$\hat{H}_{al}$, is treated clasically. The equation of motion for 
atomic operators in the interaction picture with respect to $\hat{H}_{a}$ 
are given by
\begin{eqnarray}
\label{EwoGE1}
\dot{\tilde{g}}_{\vec{n}}(t) & = & - i \frac{\Omega}{2} 
e^{i \Delta t} \sum_{\vec{m}} \tilde{e}_{\vec{m}}(t)
e^{i(\omega_{\vec{n}}-\omega_{\vec{m}})t} \langle \vec{n} | 
e^{-i \vec{k}_{L} \vec{r}} | \vec{m} \rangle,  \\
\label{EwoGE2}
\dot{\tilde{e}}_{\vec{m}}(t) & = & - i \frac{\Omega}{2} 
e^{-i \Delta t} \sum_{\vec{n}} \tilde{g}_{\vec{n}}(t)
e^{i(\omega_{\vec{m}}-\omega_{\vec{n}})t} \langle \vec{m} | 
e^{i \vec{k}_{L} \vec{r}} | \vec{n} \rangle  ,
\end{eqnarray}
where $\Delta = \omega_{L} - \omega_{0}$ is the detuning of the laser.
Now the assumption of the short pulse can be invoked to approximate 
the factors: $e^{i(\omega_{\vec{m}}-\omega_{\vec{n}})t} \approx 1$. Then, the
system of equations (\ref{EwoGE1})-(\ref{EwoGE2}) is easy to solve
analytically. To this end we introduce the operator
$\tilde{E}_{\vec{n}}(t) = \sum_{\vec{m}} 
\langle \vec{n} | e^{-i \vec{k}_{L} \vec{r}} | \vec{m} \rangle
\tilde{e}_{\vec{m}}(t)$.
The system of infitely many coupled equations then reduces to a set with
only pairwise coupling. In the next step, we insert the found solution into 
the evolution equation for operator $\hat{a}_{\vec{k} \lambda}$. In our
approach the atom charges and currents distribution is a source for the vacuum 
modes of the e-m field, while the emmitted photons have no influence on 
the evolution of atoms. In other words we neglect the interaction of scattered 
photons with other atoms. The equation of motion is solved by simple 
integration in time, which leads to 
\begin{eqnarray}
\label{SolA}
\hat{a}_{\vec{k} \lambda} (t) & = & e^{- i \omega_{\vec{k}} t} \left[ 
\hat{a}_{\vec{k} \lambda} + 
\sqrt{\frac{\omega_{\vec{k}}}{2 \hbar \varepsilon_{0} (2\pi)^3}} 
(\vec{d} \cdot \vec{\epsilon}_{\vec{k} \lambda}) 
\frac{\Omega}{{\Omega^{\prime}}^2} \right. \nonumber \\
& \times & e^{ i (\omega_{\vec{k}}-\omega_{L}) t / 2}
\left( f_{1}(t) \sum_{\vec{n},\vec{m}} \hat{g}_{\vec{n}}^{\dag} 
\langle \vec{n}| e^{i (\vec{k}_{L}-\vec{k}) \vec{r}} | \vec{m} \rangle
\hat{g}_{\vec{m}} \right. \nonumber \\
& + & \left. \left. f_{2}(t) 
\sum_{\vec{n},\vec{m}} \hat{e}_{\vec{m}}^{\dag}
\langle \vec{m}| e^{i (2\vec{k}_{L}-\vec{k}) \vec{r}} | \vec{n} \rangle
\hat{g}_{\vec{n}} + \ldots
\right) \right]    ,
\end{eqnarray}
where $\Omega^{\prime} = \sqrt{ \Delta^{2} + \Omega^{2}}$, and $f_{1}(t)$,
$f_{2}(t)$ are functions of time given in Appendix A. All the operators on
the RHS of Eq. (\ref{SolA}) without specified depedence on time 
are taken at the time $t = 0$. We will use this convention in the next 
sections. The dots in the brackets denote the remaining terms proportional
to $\hat{g}_{\vec{n}}^{\dag} \hat{e}_{\vec{m}}$ and
$\hat{e}_{\vec{m}}^{\dag} \hat{e}_{\vec{m}^{\prime}}$. They do not give any
contribution to the mean number and variance of photons.   

\section{Mean number of scattered photons}

The basic statistical quantity we can construct from the solution
(\ref{SolA}) is the mean number of photons. We would like to know if it 
depends on the statistical properties of the condensate. In particular we
are interested in its relation to the condensate fluctuations $\delta^2 N_0$.
In an experiment the angular distribution of scattered photons may be
measured by scattering the series of short pulses, then calculating the
mean from a number of detected photons in every pulse at the given angle. The
calculated value will be approximately the mean number of photons of given
mode, summed over polarization and integrated over the frequencies:
\begin{equation}
\label{Nph}
\frac{d N_{ph}}{d \Omega}(\theta,\phi,t) = \sum_{\lambda} \int \! \! \! 
dk \ k^2
\langle \hat{a}_{\vec{k} \lambda}^{\dag}(t) \hat{a}_{\vec{k} \lambda}(t)
\rangle.
\end{equation}
The $\vec{k}$ under the mean value on the RHS of Eq (\ref{Nph}) has a direction
given by $(\theta,\phi)$. As we show later the spectrum has two spectral
components, and we do the integration in Eq. (\ref{Nph}) for every component
separately. Hence, the integration leads to the total intensity of a given
peak in the spectrum. The mean value under the integral should be calculated
by tracing with the particular density
matrix. Since we work in the Heisenberg picture, this density matrix
should be the initial one for the whole system: atoms and e-m field. 
At $t=0$ the
density matrix is just the product of the density matrices of every
subsystem: $\rho(0) = \rho_g(0) \otimes \rho_e(0) \otimes \rho_f(0)$, where
letters $g$, $e$, $f$ correspond to ground state atoms, excited atoms, and
E-M field respectively. We can assume that for $t=0$ all atoms were not
excited, hence $\rho_e(0) = |0,0,\ldots \rangle \langle 0,0,\ldots |$  

Our approach pertains to a so called weak condensation regime, 
that is the situation in which the number of atoms in 
the trap is not too large, and the fluctuations are 
therefore relatively more important. In this situation 
the role of interatomic interactions is to assure 
the migration of the system in the phase space, so that 
the density matrix can be well described by a density matrix of an ideal gas 
in microcanonical, or alternatively canonical ensemble.
Experimental conditions favour the microcanonical description of
the condensate (energy and the number of particles conserved), therefore the
density matrix $\rho_g(0)$ is determined by the microcanonical
ensemble \cite{ideal}. At 
the initial moment the only occupied modes of the e-m field are these
corresponding to the laser light. But the occupation of these particular
modes does not affect the results of the measurements that are done 
for the nonzero angles. Therefore we  put 
$\rho_f(0) = |0,0,\ldots \rangle \langle 0,0,\ldots |$ 
at the initial moment.

Substituting the annihilation and creation operators in Eq. (\ref{Nph}) by
explicit formula (\ref{SolA}), after some calculations (see details in
Appendix A) we obtain
\begin{equation}
\label{SredFot}
\frac{d N_{ph}}{d \Omega \ d t}(\theta,\phi) =  \frac{d^2 \Omega^2}
{32 \pi^2 \epsilon_0 \hbar c^3 \Delta^2} \left(1 - (\vec{n}_k \vec{n}_d)^2
\right) {\omega}^3 f(\vec{q}),
\end{equation}
and 
\begin{equation}
\label{DefF}
f(\vec{q}) = \sum_{ \vec{n},\vec{n}^{\prime},\vec{m},\vec{m}^{\prime}}
\langle \hat{g}_{\vec{n}}^{\dag} \hat{g}_{\vec{n}^{\prime}} 
\hat{g}_{\vec{m}}^{\dag} \hat{g}_{\vec{m}^{\prime}} \rangle
\langle \vec{n}| e^{i \vec{q} \vec{r}} | \vec{n}^{\prime} \rangle
\langle \vec{m}| e^{- i \vec{q} \vec{r}} | \vec{m}^{\prime} \rangle    .
\end{equation}     
where $\vec{n}_k$ is the unit wector in the direction $(\theta,\phi)$ and
$\vec{n}_d$ is the unit wector in the direction of the dipole moment
$\vec{d}$. The frequency $\omega$ is the frequency of the given component of the
spectrum. Following the calculation in Appendix A, the spectrum consists of two 
peaks with frequencies: $\omega_0$ (inelastic component)
 and $\omega_L$ (elastic component). The vector $\vec{q}$ is
the vector of the momentum transfer: $\vec{q} = \vec{k} - \vec{k_L}$. 
The intensities of both elastic and inelastic components are almost the same. 
Note that in the presence of spontaneous emission losses the inelastic peak
would be suppresed for longer pulses.

For the numerical purposes our expressions (\ref{DefF}) were
modified. After some calculations the mean value involving the operators 
$\hat{g}$ may be replaced by the statistical quantities for the 
microcanonical ensemble (see Eq. (\ref{Fdlugi}) in Appendix B).
The statistical moments can be expressed by the microcanonical partition 
function, and then calculated by means of the recurrence algorithms.  
In order to check if the influence of the fluctuations may be noticed in
the measurments of scattered photons, we compare the microcanonical results
to the scattering on the nonfluctuating condensate. Such state may be
theoretically realized in the following way: the mean occupation number for
each level is the same as for the microcanonical condensate, while the
higher statistical moments are decorrelated
\begin{equation}
\label{PropRozk}
\langle N_i N_j \rangle  =  \langle N_i \rangle \langle N_j \rangle    ,
\end{equation}
where indices $i$ and $j$ label the states in the trap\footnote{ We omit the
vector indicies because the mean ooccupation number depends only on the
energy of each state and not on its spatial properties. Therefore the energy
is sufficient to enumerate the states.}. From this property it directly
follows that the condensate does not fluctuate: $\delta N_0 = 0$ in 
this particular state.

There are other possible statistical properties of the condensate. The
time-honored Bogolyubov approximation, taken at its face-value, assumes that
the condensate is in a coherent state \cite{Bogol}. We therefore consider also
the Poissonian distribution: 
\begin{eqnarray}
\label{PropKoh}
\langle N_i^2 \rangle & = & \langle N_i \rangle^2 + \langle N_i \rangle ,\\
\langle N_i N_j \rangle & = & \langle N_i \rangle \langle N_j \rangle, 
\ \ i \neq j.
\end{eqnarray}
The assumption that excited states are also coherent may be questionable,
but we do the calculations for low 
temperatures, where the statistical properties of the condensate are
dominant. For completness of our review of the
different states we included also the results for the condensate
described by thermal state. 
Of course this statistics, implicit in the grand canonical ensemble has
already been ruled-out by the experiment \cite{jila23}. The statistical properties of this 
last state are imposed by
\begin{eqnarray}
\label{PropTerm} 
\langle N_i^2 \rangle & = & 2 \langle N_i \rangle^2 + \langle N_i \rangle, 
\\     
\langle N_i N_j \rangle & = & \langle N_i \rangle \langle N_j \rangle,
\ \ i \neq j.
\end{eqnarray}
The comparison of the scattering of photons on these different condensates
consisting of 1000 atoms is presented in Fig.1. 
We plot here the elastic part of the scattering. The conclusions for the
inelastic component are very similar.
The quantity proportional to the mean number of
photons is plotted as a function of the angle $\theta$. 
This mean photon falls-off exponentially for large angles due to the presence 
of the, so called, Debye factor (see Appendix A).
The results were computed
for realistic parameters. The oscilator length $\xi$ was calculated for the 
sodium atoms in the spherically symmetric trap with $\omega = 400 s^{-1}$ . 
The resonance frequency of the transition $\omega_0$ corresponds to the 
wavelength $\lambda_0 = 800$nm, and detuning 
$\Delta =1$GHz. Analysis of Fig.1 leads to the conclusion, that only
scattering on the "thermal" condensate may be clearly distuingished. There are,
of course, differences between the scattering on the remaining three
models, but they are small, and become even smaller if the
number of atoms increases. Our main conclusion in this section is that the 
mean values do not allow to detect and measure the condensate fluctuations.      

\begin{figure}[tb]
\centering
\includegraphics[angle=90,width=8.5cm,clip]{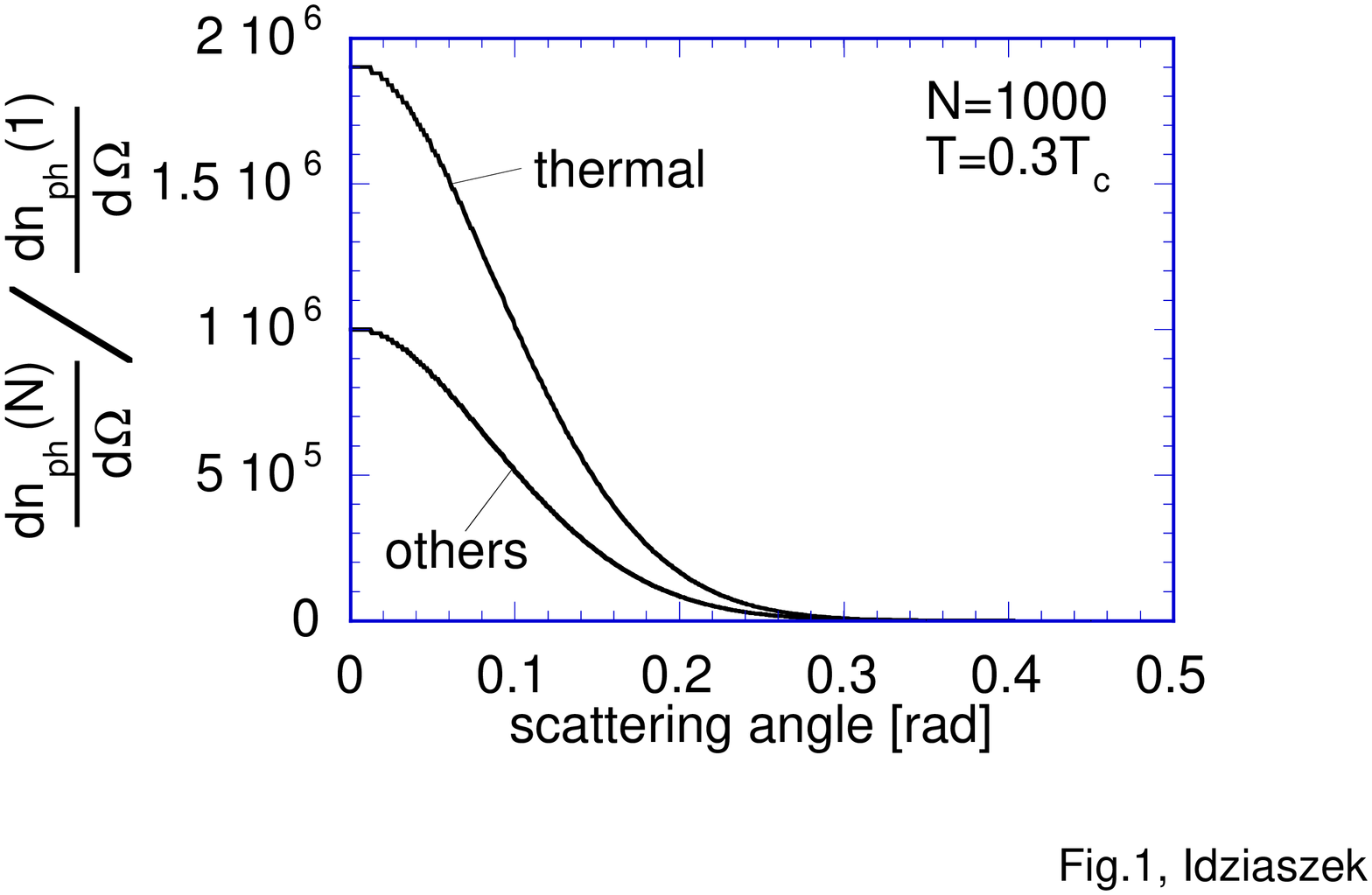}
\caption{Angular distribution of the mean number of the scattered photons
measured with respect to the direction of the incident laser pulse in the
plane perpendicular to the polarization axis of the incident light. The
elastic component of the spectrum is shown. Of four different density
matrices described in the text only the result for the thermally fluctuating
source may be distinguished. The mean number of photons is scaled by the
corresponding results for a single atom.}
\label{F1} 
\end{figure}

\section{The variance}

As we have seen above, the mean number of scattered photons is fairly
insensitive to the statistics of the condensate. We have to look at higher
moments of the photon statistics. Its variance is a suitable quantity.
It may be experimetally measured. It is sufficient only to
scatter a series of pulses and after that to calculate the variance of the 
number of photons detected during every pulse at given angle. Repeating the 
series of measurements for different angles leads to the angular dependence of
the variance. The formula which corresponds to this measured quantity is the
following
\begin{eqnarray}
\label{dNPh}
\delta \! \left( \! \frac{d N_{ph}}{d \Omega} \! \right) \! (\theta,\phi,t) &=& 
\left[ \left\langle \left[ 
\sum_{\lambda} \! \int \! \! \! dk \ k^2 
\hat{a}_{\vec{k} \lambda}^{\dag}(t) 
\hat{a}_{\vec{k} \lambda}(t) \right]^2 \right\rangle \right. \nonumber \\
&-& \left. \left\langle 
\sum_{\lambda} \! \int \! \! \! dk \ k^2 
\hat{a}_{\vec{k} \lambda}^{\dag}(t)
\hat{a}_{\vec{k} \lambda}(t) \right\rangle^2 \right]^{\frac{1}{2}} \! \!.  
\end{eqnarray}
The mean values under the integrals should be calculated as in the previous
case by tracing with density matrix of the initial state. The integration
over the frequency is performed for each constituent of the spectrum
separately, because we are interested in checking the
variance of the photons both in elastic and inelastic scattering.
Substituting the solution (\ref{SolA}) for the creation and annihilation 
operators in (\ref{dNPh}), 
after a tedious but straightforward calculations we
obtain the following result
\begin{equation}
\label{WarFot}
\delta \! \left( \frac{d N_{ph}}{d\Omega dt} \right) \! (\theta,\phi) =  
\frac{d^2 \Omega^2 \omega^3}{32 \pi^2 \epsilon_0 \hbar c^3 \Delta^2}
\left(1 - (\vec{n}_k \vec{n}_d)^2
\right)  \sqrt{g(\vec{q})}    ,
\end{equation}
and
\begin{eqnarray}
\label{DefG}
g(\vec{q}) &=& \! \! \sum_{ 
\vec{n},\vec{n}^{\prime},\vec{m},\vec{m}^{\prime}
\vec{l},\vec{l}^{\prime},\vec{j},\vec{j}^{\prime} } \! \! \! \! \! \! \! \! 
\beta_{\vec{n},\vec{n}^{\prime}}(\vec{q}) 
\beta_{\vec{m},\vec{m}^{\prime}}(-\vec{q})
\beta_{\vec{l},\vec{l}^{\prime}}(\vec{q}) 
\beta_{\vec{j},\vec{j}^{\prime}}(-\vec{q}) \nonumber \\
&\times& \left( \langle \hat{g}_{\vec{n}}^{\dag} \hat{g}_{\vec{n}^{\prime}}
\hat{g}_{\vec{m}}^{\dag} \hat{g}_{\vec{m}^{\prime}} 
\hat{g}_{\vec{l}}^{\dag} \hat{g}_{\vec{l}^{\prime}}
\hat{g}_{\vec{j}}^{\dag} \hat{g}_{\vec{j}^{\prime}} 
\rangle \right. \nonumber \\
&-& \left. \langle \hat{g}_{\vec{n}}^{\dag} \hat{g}_{\vec{n}^{\prime}}
\hat{g}_{\vec{m}}^{\dag} \hat{g}_{\vec{m}^{\prime}} \rangle
\langle \hat{g}_{\vec{l}}^{\dag} \hat{g}_{\vec{l}^{\prime}}
\hat{g}_{\vec{j}}^{\dag} \hat{g}_{\vec{j}^{\prime}} \rangle \right),
\end{eqnarray}
where we denote Frank-Condon coefficients by
$\beta_{\vec{n},\vec{n^{\prime}}}(\vec{q}) \equiv  
\langle \vec{n}| e^{i \vec{q} \vec{r}} | \vec{n}^{\prime} \rangle$.
The Eq. (\ref{WarFot}) is very similar to Eq.(\ref{SredFot}) for the mean
value, and we use the same notation. The frequency $\omega$ is, as
previously, the frequency of the considered component of the spectrum. 
In order to perform the numerical calculations the mean values of the 
atomic operators in formula (\ref{DefG}) are expressed by means of the
different statistical moments of the occupation number of the trap states.
Unfortunately even when we couple the annihilation operators $\hat{g}$ with
the creation operators $\hat{g}^{\dag}$ into pairs, we still have to sum
over four indices. These indices are additionally the vector ones. 
Furthermore the number of possible combinations of
these pairs is much larger than in the case of the four operators in
Eq.(\ref{DefF}) for the mean number. We have to use some approximation to 
be able to perform the summation for reasonable size condensates ($\sim$ 1000 
atoms).
We may utilise the fact that for BEC the number of atoms in the ground state
is of the order of the total number of atoms, so we expand our formula
(\ref{DefG}) in the powers of $\hat{g}_{0}^{\dag} \hat{g}_{0}$. 
As it turns out, for 1000 atoms a sufficient approximation requires inclusion  
of fourth-, third- and second order terms in this expansion. In this case 
summation is performed over two different excited states, what is possible 
to realize.

In Fig.2  we present the variance of the scattered photons calculated on the 
condensates with different statistical properties, described in the previous
section. The
numerical computations were realized for the same physical parameters. 
Again we only present the results for the elastic component.
For vertical axis we also choose the same units as in the previous picture.
Now we are able to 
distinguish the scattering on the coherent state, what was not possible on
the basis of mean numbers. The larger difference occures for the elastic 
scattering for small angles. The difference between scattering on
the microcanonical and uncorrelated condensated is still very small.
However, for small angles these curves are different.
In this regime our approximation is not valid for the microcanonical 
condensate.
The expansion of the variance in the powers of $q$, shows that the
result behaves as $\theta^2$ and the curve is indistinguishable from the 
curve for uncorrelated condensate.        
As we expected, the scattering on "thermal" condensate leads to
completely different results. In fact while microcanonical and uncorrelated
states are even harder to distinguish for larger number of atoms, the
distance to the results for the coherent states grows linearly with $N$.  

\begin{figure}
\centering
\includegraphics[width=8.5cm,clip]{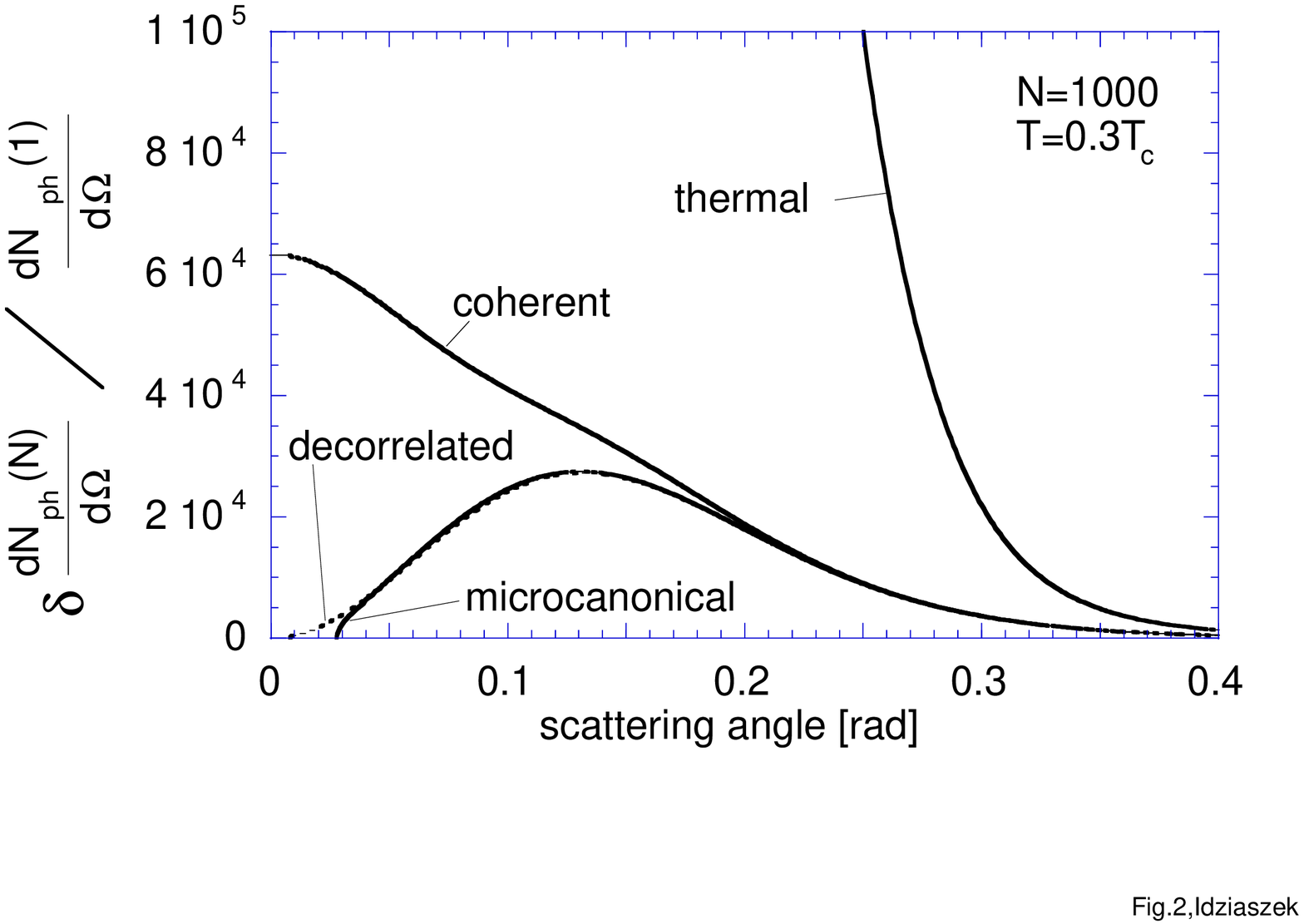}
\caption{Angular distribution of the variance of the number of scattered
photons is shown. As in Fig.\ref{F1} the elastic component of the spectrum is
shown. Note that now not only the thermal but also the coherent states
results are clearly distinguish. It is still impossible, however, to
distinguish between the microcanonical and the decorrelated results.}
\label{F2}
\end{figure}

\section{Conclusions}

The statistical properties of the Bose-Einstein condensate are of great 
interest. Its measurments would advance the theory of the coherence of 
existing atom lasers.

In the present paper we propose to use the scattering of short laser pulses
for the study of fluctuations. One needs higher moments of the distribution
of the number of the scattered photons to have a sufficiently subtle tool.

The main open question is the influence of weak interactions on the statistics.
While the theory remains muddled, we know that the main impact of interactions
is the emergence of two length-scales: that of the size of the condensate,
broadened by the repulsive forces and that of the size of thermal cloud, 
expelled from the center of the trap by the condensate. It means that we would 
have two 
different Debye factors damping the scattering cross-section at large angles.
This way by changing the angle one could, perhaps, gain access to the 
fluctuations of the condensate and fluctuations of the thermal cloud 
separately. 
They are lumped together for the ideal Bose gas studied in the present paper.

\section{Acknowledgments}
We would like to thank M. Gajda for fruitful discussions.
The idea of employing photon statistics for measuring statistical properties
of condensate was suggested by J. I. Cirac, 
M. Lewenstein and L. You, and presented at 
the Cold Atom Workshop at ITAMP at Harvard University and
Smithsonian Astrophysical Observatory.  
Z.I. and K. R were supported by the 
{\it Polish KBN} Grant 203 B 05715 and the {\it Subsidy 
of the Foundation for Polish Science}. 
K.R. thanks the Alexander von Humboldt-Foundation for its generous support.
This work is also supported by the SFB 407 of the {\it Deutsche 
Forschungsgemeinschaft}.

\appendix
\section{The photon spectrum}
The function of time which appears in the solution (\ref{SolA}) are given by
\begin{eqnarray}
\label{DefF1}
f_1(t) & = & 
\frac{\Delta}{\omega_k -\omega_L} \sin( \frac{\omega_k -\omega_L}{2} t)
- \frac{\Omega^{\prime} + \Delta}{4 \omega_{1k}} 
e^{\frac{i \Omega^{\prime}t}{2}} \sin(\omega_{1k} t) \nonumber \\
&+& \frac{\Omega^{\prime} - \Delta}{4 \omega_{2k}}
e^{-\frac{i \Omega^{\prime}t}{2}} \sin(\omega_{2k} t)   ,\\
f_2(t) & = &
\label{DefF2} 
\frac{\Omega }{\omega_k -\omega_L} \sin( \frac{\omega_k -\omega_L}{2} t)
- \frac{\Omega}{4 \omega_{1k}} 
e^{\frac{i \Omega^{\prime}t}{2}} \sin(\omega_{1k} t) \nonumber  \\
&-& \frac{\Omega}{4 \omega_{2k}} e^{-\frac{i \Omega^{\prime}t}{2}} 
\sin(\omega_{2k} t) ,
\end{eqnarray}
where we use the following notation: 
$\Omega^{\prime} = \sqrt{ {\Delta}^{2} + \Omega^{2}}$
$\omega_{1k} = (\omega_k -\omega_L + \Omega^{\prime})/2$,
$\omega_{2k} = (\omega_k -\omega_L - \Omega^{\prime})/2$.
Our laser pulse is weak and far detuned and we may simplify 
the equations (\ref{DefF1})-(\ref{DefF2}). For these assumptions: 
$\Delta \gg \Omega$ 
and $\Omega^{\prime} \approx \Delta$ for positive detuning.
The functions of time $f_1$ and $f_2$ take the form
\begin{eqnarray}
\label{ApprF1}
f_1(t) & = & 
\frac{\Delta}{\omega_{kL}} 
\sin \left( \frac{\omega_{kL}}{2}t \right) -
\frac{\Delta e^{i \Delta t/2}}{\omega_{k0}} 
\sin \left( \frac{\omega_{k0}}{2}t \right) ,\\
f_2(t) & = & 
\frac{\Omega}{\omega_{kL}}
\sin \left( \frac{\omega_{kL}}{2}t \right)
-\frac{\Omega e^{i \Delta t/2}}{2\omega_{k0}}  
\sin \left( \frac{\omega_{k0}}{2}t \right) \nonumber \\
&-& \frac{\Omega e^{-i \Delta t/2}}
{2(\omega_{k} + \omega_0 -2 \omega_L)} 
\sin \left( \frac{\omega_{k} + \omega_0 -2 \omega_L}{2}t \right) , 
\end{eqnarray}
where $\omega_{kL} = \omega_k - \omega_L$ and $\omega_{k0} = \omega_k -
\omega_0$.
Calculation of the mean number of photon in a particular mode is done by taking
solution (\ref{SolA}) and then calculating the mean value. This mean value 
should
be computed with the initial density matrix, whose properties are discussed
in the text. After straightforward calculations we get
\begin{eqnarray}
\label{Meanaa}
\langle \hat{a}_{\vec{k} \lambda}^{\dag}(t) \hat{a}_{\vec{k} \lambda}(t)
\rangle &=& \frac{ \omega \Omega^2(\vec{d} \cdot 
\vec{\epsilon}_{\vec{k} \lambda})^2}{16 \pi^3 \epsilon_0 \hbar \Delta^4}
\nonumber \\
&\times& \left( f(\vec{q}) |f_1(t)|^2  + N |f_2(t)|^2 \right),
\end{eqnarray}
where the function $f(\vec{q})$ is defined in Eq. (\ref{DefF}). This function 
is of
the order of $N^2$ what may be easily demonstrated. Additionally the modulus 
of the function $|f_2|$ is much smaller than $|f_1|$. Hence we are justified 
to neglect the contribution of function $|f_2|$ to our formula. In order to
compute mean number of photons we have to sum over polarizations and
integrate over frequencies. The first task may be performed with the help
of well-known identity: 
$\sum_{\lambda} (\vec{n_d} \cdot \vec{\epsilon}_{\vec{k} \lambda})^2 =
\left(1 - (\vec{n}_k \vec{n}_d)^2\right)$.
The information about the spectrum is hidden in the function $f_1(t)$. The
integration over the frequencies may be easily performed if we assume the
spectrum consist of very narrow lines. Since the pulse
duration is long in comparison to the inverse of the optical frequencies:
$t \gg \omega_L^{-1},\omega_0^{-1}$, we may approximate the spectrum to the
form:
\begin{equation}
\label{FDelta}
|f_1(t)|^2 \approx \frac{\pi \Delta^2 t}{2} \left[ \delta(\omega_{k}-\omega_0)
+ \delta(\omega_{k}-\omega_L) \right] .
\end{equation}
As one can see the spectrum consists of two peaks centered on the atomic and
laser frequencies. The intensities of the two are the same, which is the result
of the absence of the spontaneos emission during the short pulse. Now 
the result (\ref{SredFot}) is evident. 
The derivation of the Eq.(\ref{WarFot}) describing the variance  
is done in the similar way as for the mean.

\section{Numerical strategies}
The basic numerical task is to compute the functions $f(\vec{q})$ and 
$g(\vec{q})$ which enter
into equations for the mean value (\ref{SredFot}) and the variance
(\ref{WarFot}). For this purpose we express the mean value in the functions 
$f(\vec{q})$ and $g(\vec{q})$ by mean occupation numbers and higher moments 
of trap energy levels. We explain how it should be done for the
function $f(\vec{q})$, the derivation in the case of variance is similar.

The summation in the function $f(\vec{q})$
\begin{equation}
\label{DefFagain}
f(\vec{q}) = \sum_{ \vec{n},\vec{n}^{\prime},\vec{m},\vec{m}^{\prime}}
\langle \hat{g}_{\vec{n}}^{\dag} \hat{g}_{\vec{n}^{\prime}}
\hat{g}_{\vec{m}}^{\dag} \hat{g}_{\vec{m}^{\prime}} \rangle
\beta_{\vec{n},\vec{n}^{\prime}}(\vec{q}) 
\beta_{\vec{m},\vec{m}^{\prime}}(-\vec{q}),
\end{equation}
is performed over four vector indices. We may exclude two of them because
the mean of the four operators 
$\langle \hat{g}_{\vec{n}}^{\dag} \hat{g}_{\vec{n}^{\prime}}
\hat{g}_{\vec{m}}^{\dag} \hat{g}_{\vec{m}^{\prime}} \rangle$
is nonzero only if the operators couple into pairs. For four operators there
are two possibilities of pairing, and we obtain
\begin{eqnarray}
\label{Fpair}
f(\vec{q}) &=& \sum_{ \vec{n} \neq \vec{m} }
\langle \hat{N}_{\vec{n}} \hat{N}_{\vec{m}} \rangle \left(
\beta_{\vec{n},\vec{n}}(\vec{q}) \beta_{\vec{m},\vec{m}}(-\vec{q}) +
|\beta_{\vec{n},\vec{m}}(\vec{q})|^2 \right) \nonumber \\
&+& \sum_{\vec{n}} \langle \hat{N}_{\vec{n}}^2 \rangle 
\beta_{\vec{n},\vec{n}}(\vec{q}) \beta_{\vec{n},\vec{n}}(-\vec{q}) + N,
\end{eqnarray}  
where $\hat{N}_{\vec{n}} \equiv \hat{g}_{\vec{n}}^{\dag} \hat{g}_{\vec{n}}$.
In derivation of (\ref{Fpair}) we use the identity 
$\sum_{\vec{m}} \beta_{\vec{n},\vec{m}}(\vec{q})
\beta_{\vec{m},\vec{n}^{\prime}}(-\vec{q}) =
\delta_{\vec{n},\vec{n}^{\prime}}$ that results directly from the
completeness of the states. The Franck-Condon coefficients for the harmonic
trap may be calculated analytically. The result for three dimensional trap
is simply the product of the 1D Franck-Condon coefficients 
\begin{equation}
\label{FrankC}
\langle n| e^{i q x}|m \rangle = \sqrt{\frac{n!}{m!}} 
e^{- q^2 \xi^2 /4} L_{n}^{m-n} \left( \frac{q^2 \xi^2}{2} \right)
\left( \frac{i q \xi}{2} \right)^{m-n},
\end{equation}
where $L_n^m(x)$ denotes Laguerre polynomial and $\xi$ is the oscillator
length. 
For simplicity we assume our
trap to be spherically symmetric, therefore we may freely choose the 
orientation 
of the system of coordinates for the states in the trap. We choose the
z-axis to be in the direction of the $\vec{q}$ vector. For such system of
coordinates three dimensional Franck-Condon coefficients become 
$\langle \vec{n}| e^{i \vec{q}\vec{r}}|\vec{m} \rangle =
\langle n_z| e^{i q z}|m_z \rangle \delta_{n_x,m_x} \delta_{n_y,m_y}$. 
Since the mean values of the different combinations of $\hat{N}_{\vec{n}}$
operators depends only on the energies of the states, and not on the
distribution of the quantum numbers, we may split the summation into two
steps: the sum over energies and the sum over degeneracies. 
The last one may be performed for all
terms, and the result is a combination of Laguerre polynomials.
For the microcanonical ensemble  
\begin{eqnarray}
\label{Fdlugi}
f(\vec{q}) & = & e^{- \eta} \! \left[ 
\langle \hat{N}_{0}^2 \rangle \! - \! \langle \hat{N}_{0} \rangle
+ 2 \! \! \sum_{E<E^{\prime}} 
\langle \hat{N}_{E} \hat{N}_{E^{\prime}} \rangle \! \left(
L_{E}^2(\eta) L_{E^{\prime}}^2(\eta) \right. \right. \nonumber \\
&+& \left. \eta^{E^{\prime}-E} S_E^{E^{\prime}-E}(\eta) \right) 
+ \sum_{E>0} \langle \hat{N}_{E} \hat{N}_{\tilde{E}} \rangle L_{E}^2(\eta)^2
\nonumber \\
&+& \! \left. \sum_{E>0} \left( \langle \hat{N}_{E}^2 \rangle \!
- \! \langle \hat{N}_{E} \rangle \!
- \! \langle \hat{N}_{E} \hat{N}_{\tilde{E}} \rangle \right) 
S_{E}^{\, 0}(\eta) \right] + N ,
\end{eqnarray}
where $\eta = q^2 \xi^2 /2$ (the exponent in front of the square bracket
is called the Debye factor) and  
the mean $\langle \hat{N}_{E} \hat{N}_{\tilde{E}} \rangle$ is the
correlation between different levels with the same energies. In
the formula (\ref{Fdlugi}) $S_m^a (x)$ denotes the sum including the Laguerre
polynomials:
\begin{equation}
\label{defS}
S_m^a (x) = \sum_{n=0}^m \frac{n!}{(n+a)!} L_n^a(x)^2 (m+1-n).
\end{equation}
The sum (\ref{defS}) may be done analytically with use of the summation 
formula 
for Laguerre polynomials 8.974.1 \cite{Sum_thm}. The calculation leads to 
\begin{eqnarray}
\label{SumaLag}
S_m^a (x) & = &
\frac{(m \! + \! 2)!}{(m \! + \! a)!} \! \left( 
L_m^{a+1} L_m^{a} \! - \! L_{m-1}^{a+1} L_{m+1}^{a} \right)
-  \frac{(m \! + \! 1)!}{(m \! + \! a \! - \! 1)!} \nonumber \\  
&\times& \left[ \frac{1}{6} L_{m-2}^{a+2} L_{m}^{a-1} -
\frac{1}{6} L_{m-3}^{a+2} L_{m+1}^{a-1}  - \frac{1}{2} L_{m-2}^{a+1} L_{m}^{a}
\right. \nonumber \\
 &+& \left.
\frac{1}{2} L_{m-1}^{a+1} L_{m-1}^{a} \right]  - \frac{(m \! + \! 1)!}
{(m \! + \! a)!} (L_m^a)^2,
\end{eqnarray}
where we use shorthand notation $L_{m}^{a} \equiv L_{m}^{a}(x)$ and assume 
$L_m^a(x)=0$ for $m<0$.

\end{document}